\documentclass[manuscript, screen]{acmart}
\usepackage{latexsym}
\usepackage{graphicx}
\usepackage{amsmath}
\usepackage{array}
\usepackage{color}
\usepackage[hang, flushmargin]{footmisc}
\usepackage{url}

\AtBeginDocument{%
  \providecommand\BibTeX{{%
    \normalfont B\kern-0.5em{\scshape i\kern-0.25em b}\kern-0.8em\TeX}}}
    
\newcommand\BibTeX{B{\sc ib}\TeX}

\copyrightyear{2021}
\acmDOI{10.1145/3430803}

\begin{document}

\title{Gender Trends in Computer Science Authorship}


\author{Lucy Lu Wang}
\email{lucyw@allenai.org}
\orcid{0000-0001-8752-6635}
\affiliation{
  \institution{Allen Institute for Artificial Intelligence}
  \city{Seattle}
  \state{Washington}
  \postcode{98103}
}

\author{Gabriel Stanovsky}
\thanks{G. Stanovsky: Work done while at the Allen Institute for Artificial Intelligence and the University of Washington.}
\email{gabis@allenai.org}
\affiliation{
  \institution{The Hebrew University of Jerusalem}
  \state{Israel}
}

\author{Luca Weihs}
\email{lucaw@allenai.org}
\affiliation{
  \institution{Allen Institute for Artificial Intelligence}
  \city{Seattle}
  \state{Washington}
  \postcode{98103}
}

\author{Oren Etzioni}
\email{orene@allenai.org}
\affiliation{
  \institution{Allen Institute for Artificial Intelligence}
  \city{Seattle}
  \state{Washington}
  \postcode{98103}
}

\renewcommand{\shortauthors}{Wang LL et al.}

\begin{abstract}
  A large-scale, up-to-date analysis of Computer Science literature (11.8M papers through 2019) reveals that, if trends from the last 50 years continue, parity between the number of male and female authors will not be reached in this century. In contrast, parity is projected to be reached within two to three decades or may have already been reached in other fields of study like Medicine or Sociology. Our analysis of collaboration trends in Computer Science reveals shifts in the size of the collaboration gap between authors of different perceived genders. The gap is persistent but shrinking, corresponding to a slow increase in the rate of cross-gender collaborations over time. Together, these trends describe a persistent gender gap in the authorship of Computer Science literature that may not close without systematic intervention.
\end{abstract}





% teaser figure
\begin{teaserfigure}
\centering
\begin{minipage}[t]{.48\textwidth}
  \centering
  \includegraphics[width=\linewidth,keepaspectratio]{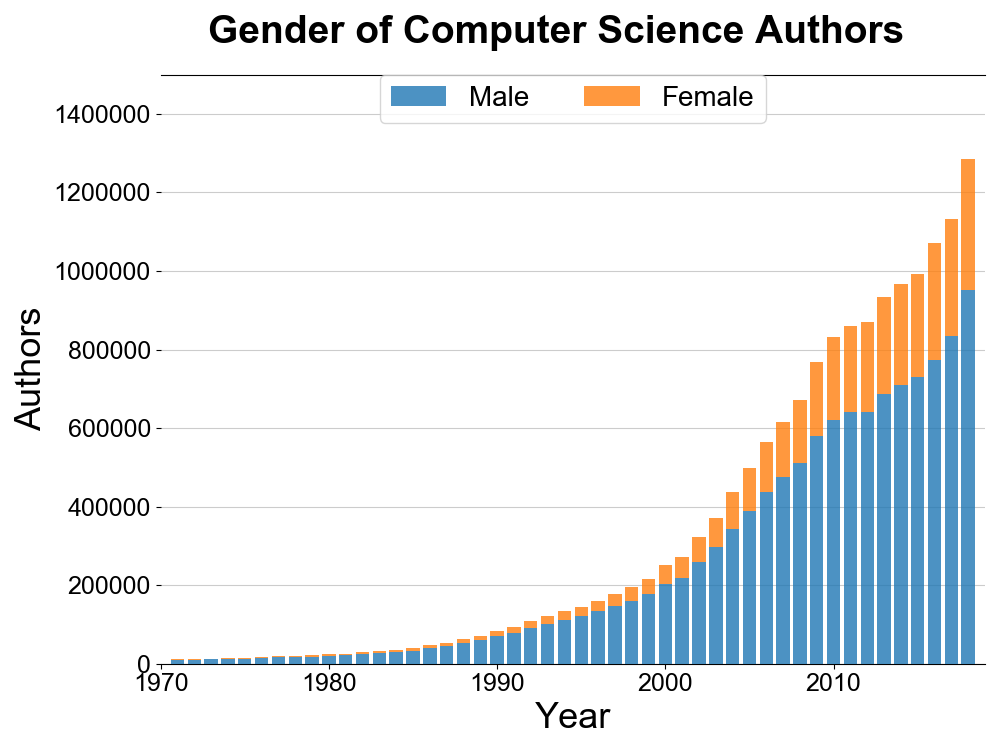}
  \captionof{figure}{\label{fig:cs_authors} Gender of Computer Science authors over time, computed by averaging across gender probabilities in our dataset.}
\end{minipage}%
\hspace{4mm}
\begin{minipage}[t]{.48\textwidth}
  \centering
  \includegraphics[width=\linewidth,keepaspectratio]{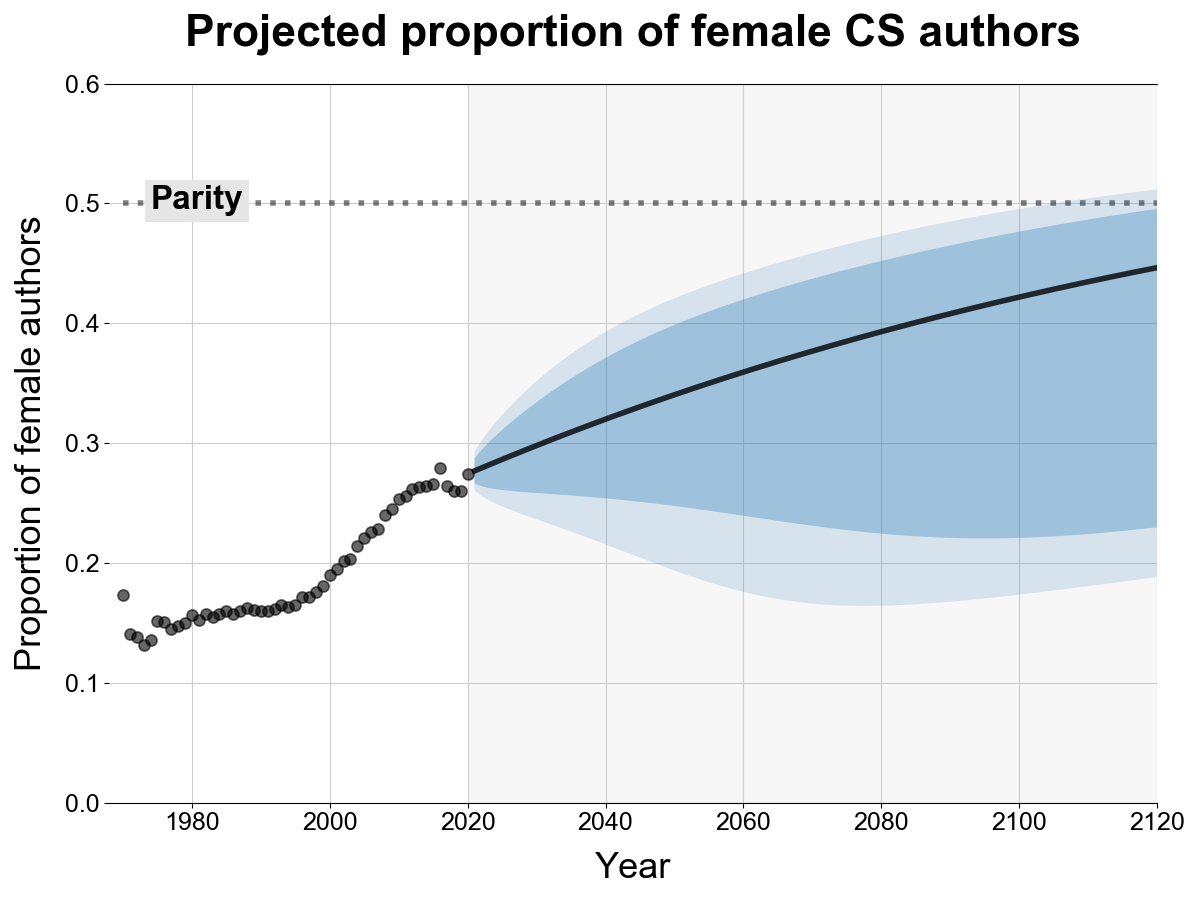}
  \captionof{figure}{\label{fig:cs_trends} The proportion of female authors is projected using an ARIMA model assuming logistic growth towards equilibrium proportions in the range [0.3, 1.0]. Confidence intervals at 80\% and 95\% are shown.}
\end{minipage}
\end{teaserfigure}

\maketitle

\section{Introduction}

This paper presents a large-scale automated analysis of gender trends in the authorship of Computer Science literature. Specifically, we aim to address the following questions:

\begin{itemize}
    \item How is gender balance among authors changing over time?
    \item When might gender parity be reached among authors?
    \item How is gender associated with co-authorship?
    \item How does Computer Science compare to other fields of study in gender representation among authors?
\end{itemize}

We answer these questions by performing an automated study of literature metadata from scientific conferences and journals, using data from the Semantic Scholar academic search engine.\footnote{\url{https://www.semanticscholar.org/}} Our study incorporates metadata from 11.8M Computer Science publications. To provide a basis for comparison, we also analyze more than 140M papers from other fields of study. We attempt to provide an overview of the relationship between gender and authorship in Computer Science, both throughout the history of the field, as well as in relation to other fields of study. Our results demonstrate that although progress has been made, there is still a significant gap in gender representation among Computer Science authors. Continued delay in addressing the gender gap may perpetuate imbalances for generations to come.

\section{Data}

\begin{table*}[tb!]
    \centering
    \small
    \begin{tabular}{p{35mm}>{\centering\arraybackslash}p{22mm}>{\centering\arraybackslash}p{22mm}>{\centering\arraybackslash}p{22mm}}
        \toprule
         Field of Study & Total papers & Total author- paper units & Average authors per paper \\
        \midrule
         Art & 5.3M & 7.4M & 1.4 \\
         Biology & 15.1M & 55.2M & 3.7 \\
         Business & 3.7M & 5.8M & 1.6 \\
         Chemistry & 14.7M & 48.6M & 3.3 \\
         \textbf{Computer Science} & \textbf{11.8M} & \textbf{27.3M} & \textbf{2.3} \\
         Economics & 3.8M & 6.4M & 1.7 \\
         Engineering & 10.1M & 20.9M & 2.1 \\
         Environmental Science & 2.0M & 4.6M & 2.3 \\
         Geography & 4.0M & 7.3M & 1.8 \\
         Geology & 3.2M & 8.4M & 2.6 \\
         History & 6.0M & 8.2M & 1.4 \\
         Materials Science & 7.4M & 21.7M & 2.9 \\
         Mathematics & 5.5M & 10.9M & 2.0 \\
         Medicine & 32.4M & 111.9M & 3.4 \\
         Philosophy & 2.8M & 3.9M & 1.4 \\
         Physics & 7.8M & 31.0M & 4.0 \\
         Political Science & 4.9M & 6.8M & 1.4 \\
         Psychology & 7.0M & 14.7M & 2.1 \\
         Sociology & 4.6M & 6.3M & 1.4 \\
        \midrule
         Total & 152.1M & 407.2M & 2.7 \\
        \bottomrule
    \end{tabular}
    \caption{Corpus statistics for different fields of study.}
    \label{tab:summary_stats}
\end{table*}

Our analysis was performed over the Semantic Scholar literature corpus \citep{Ammar2018ConstructionOT}. The corpus contains publications from between 1940 and the end of November 2019, and associated metadata such as title, abstract, authors, publication venue, and year of publication. Metadata in Semantic Scholar are derived from academic publishers, as well as scientific repositories like arXiv, DBLP, and PubMed. We use the 19 fields of study defined by Microsoft Academic \citep{shen-etal-2018-web}, which are integrated with Semantic Scholar data. Table \ref{tab:summary_stats} shows the distribution of papers used in our analysis by field of study.

The author list is extracted from all publications and compiled into a list of first names. We use Gender API\footnote{\url{https://gender-api.com/}} to perform gender lookup for each name. Gender API is a large online database of known name-gender relationships derived by linking publicly available governmental data with social media profiles in various countries. For each name, Gender API outputs the predicted binary gender (\emph{female} or \emph{male}), along with the accuracy associated with the prediction and the number of samples used to arrive at that determination. We exclude authors for whom first names are missing, and for whom only first initials are available. We also filter out first names that occur less than 10 times in our overall corpus, to reduce the number of API calls to manageable numbers.

Because many names are ambiguous with respect to gender, we use the accuracy returned by Gender API to represent the gender of each author as a distribution over male and female probabilities. For example, Gender API estimates the first name Matthew to be male with an accuracy score of 100, the maximum. The name Taylor, however, is estimated to be female but only receives an accuracy score of 55. These accuracies are used to generate two probabilities for each name, $(m, f)$, where $m$ is the probability of the associated author being perceived as male, and $f$ is the probability of the associated author being perceived as female, where $m+f=1$. In this example, each author named Matthew will be represented with the probability tuple $(1.0, 0.0)$, and each author named Taylor will be represented as $(0.45, 0.55)$.

We acknowledge that gender identity is fluid and non-binary. However, for the sake of this large-scale study--we adopt a simplified view of gender as a probability distribution over two genders, relying on first names as a proxy for the author's perceived gender (as opposed to self-reported gender). We use Gender API's results as an estimation of authors' perceived binary gender, and use these estimates to generalize over our corpus. We are not making claims about any author's true self-reported gender.

\section{Analyses}
We perform two types of analysis on this data. First, we analyze publication trends, examining the number and proportion of female authors over time (\S \ref{sec:authorship}). To identify when gender parity may be reached, we project the proportion of female authors based on trends from the last 50 years (since 1970).
In this paper, we define parity as the proportion of female authors falling within 10\% of 0.5, within the range of 0.45-0.55. We also study trends in co-authorship behavior as reflected in our data (\S \ref{sec:collab}).

\subsection{Authorship analysis}
\label{sec:authorship}

Most papers are authored by more than one individual. For the purposes of our analysis, each author-paper pair is treated as one unit. A paper with a single author yields one author-paper unit; a paper with three authors yields three author-paper units and so on. In Computer Science , the average number of authors is approximately 2.3 per paper. However, average authors per paper have increased from approximately 1.5 per paper in 1970 to approximately 3.0 in the past several years, which reflects patterns observed by other researchers \citep{Fernandes2016EvolutionIT}. Appendix \ref{app:app_over_time} provides further discussion of this shift in relation to concurrent increases in author count in other fields.

The proportion of female authors over time is used to project the trend towards gender parity. The number of female authors in a given year is computed as the sum of probabilities $f$ over the author-paper units of that year, and the number of male authors is correspondingly generated as the sum of probabilities $m$. The proportion of female authors for each year $F_t$ is computed as the number of female author-paper units divided by the total number of author-paper units for the corresponding year. We compute projections by performing an autoregressive integrated moving average (ARIMA) analysis, a widely used and established method for creating time series forecasting models \citep{TimeS1994Box}. ARIMA is an autoregressive forecasting technique: which means it uses historical values in a time series to predict current and future values. We use the auto ARIMA function in the R `forecast' package \citep{Automatic2008Hyndman}, which automates the selection of ARIMA model order, with a preference for simple models with lower order.

We assume that the growth in female author proportion observes logistic behavior. The proportion of female authors is necessarily constrained between 0 and 1, and logistic growth assumes that a stable equilibrium will eventually be reached. We tested other fit functions (linear and exponential; see Appendix \ref{app:fit_rmse} for details), but found them to be less suitable; the root-mean-squared-error (RMSE) of the logistic fit is lower than that of these other curve types when fitting to the growth curves of each field of study.

To perform the fit, we first apply $\sigma_{\alpha}^{-1}$, the inverse of the $\alpha$-scaled sigmoid (or logit) function $\sigma_\alpha(x) = \alpha / (1 + \exp(-x))$, to map the gender proportion into the real line so that the data is more amenable to linear approximation. We call $\alpha$ the expected equilibrium proportion parameter. This transform generates $y_t = \sigma_\alpha^{-1}(F_t)$, where $F_t$ is the proportion of female authors per year. We then fit a non-seasonal ARIMA model with parameters $p$, $d$, and $q$ for the transformed process $y_t$ represented by the following equation:
\begin{equation}
\label{eq:1}
    \phi_p(B)(1-B^d)y_t = c + \theta_q(B)\varepsilon_t
\end{equation}

\noindent where $B$ is the backshift operator, which shifts by one to the previous time point, and $\varepsilon_t$ is zero-centered, normally distributed noise \citep{Automatic2008Hyndman}.

Finally, we obtain the forecast in the original domain using a sigmoid transform over the projected values, applying $\sigma_\alpha$ to $y_t$ for $t > 2019$. We sample $\alpha$ from the range $[0.3, 1.0]$ so that $\sigma_{\alpha}$ has minimum and maximum values of 0 and $\alpha$ respectively. This constrains the projected values to be between 0 and some expected equilibrium proportion defined by $\alpha$. The 80\% and 95\% confidence intervals of the prediction are computed from averaging the projection confidence over 10000 iterations of model fitting.

The range for $\alpha$ is defined based on the space of likely equilibrium proportions, as estimated based on trends observed in various fields of study (see Figure \ref{fig:prop_by_fos}). Note that $\alpha$ represents the proportion of female authors we expect in the long run. An equilibrium proportion of 0.5 indicates that we expect the authorship makeup to eventually stabilize at around 50\% men and 50\% women. An equilibrium proportion of 0.9 indicates that we expect the authorship makeup to eventually stabilize at around 10\% men and 90\% women. As is further elaborated in \S \ref{sec:authorship-trends}, we perform a sensitivity analysis to determine the effect of the selected $\alpha$ parameter on the year in which parity is expected to be reached.

\subsection{Co-authorship analysis}
\label{sec:collab}
Co-authorship is computed for each unique pair of author-paper pairs for each paper. If a paper has $n$ authors, $n \choose 2$ co-author pairs are generated. Given one co-author pair $(n_1, n_2)$ and associated gender probabilities $n_1 \rightarrow (m_1, f_1)$ and $n_2 \rightarrow (m_2, f_2)$, we compute three probabilities, $p_{mm}$, $p_{mf}$, and $p_{ff}$, corresponding to the gender combinations, i.e., 
between two male authors, a male and a female author, and two female authors respectively:
\begin{equation}
    p_{mm} = m_1 m_2; \quad
    p_{mf} = m_1 f_2 + f_1 m_2; \quad
    p_{ff} = f_1 f_2
\end{equation}

\noindent where \(p_{mm} + p_{mf} + p_{ff} = 1\). The numbers of male-male, male-female, and female-female co-author pairs for each year are computed by summing over the above probabilities over all co-authorship pairs of that year. 

We then assess the number of same-gender and different-gender collaborations over time. The results are measured as a deviation from the expected, where the expected co-authorships are determined by sampling from the numbers of female and male authors active in a given year, assuming the same number of collaborations per year as observed in our data. The total number of extra or missing collaborations is computed as the difference between the observed counts of each type of collaboration and the expected value. To show rates of change, we also compute the ratio between observed and expected collaborations (O/E) of each type.

\section{Results}

In the following section, we discuss the main findings of our study.

\subsection{Gender API results} \label{sec:gender-results}

The 152.1M papers in our corpus resulted in 407.2M author-paper units. Of these author units, 14.5M lack first names, 110.0M have only a first initial, and 5.7M have a first name that occurs less than 10 times in the corpus. These author units are removed from further analysis. The remaining 277.0M author units are associated with 521K unique first names. We query these 521K names in Gender API, and acquire gender information for 351K; 170K names have insufficient information and are excluded from analysis. Of the 11.8M papers in Computer Science and the 27.3M author-paper units therein, 24.1M authors have valid first names, and 16.9M author-paper units (61.8\%) resulted in associated gender information, which is higher coverage compared to authors in other fields (we acquire gender information for approximately 50.4\% of authors across all fields).

\subsection{Gender trends among authors} \label{sec:authorship-trends}

Figure \ref{fig:cs_authors} shows that the overall author count in Computer Science has increased substantially over the last several decades, as the field has experienced significant growth. The total number of author-paper units in 2018 is above 1.2M. The proportion of female authors has also increased during this time.

Figure \ref{fig:cs_trends} shows the projected proportion of female authors in Computer Science. The projected growth in female author proportion is computed using ARIMA. We assume logistic growth, and sample the $\alpha$ parameter for equilibrium proportion from the range $[0.3, 1.0]$. We report an average projection computed over 10000 samples. Residuals of the ARIMA fit line over the logit-transformed data appear normally distributed, and are not significant under the Shapiro-Wilk Normality Test \citep{AnA1965Shapiro}. The proportion of female authors in Computer Science is predicted to reach 0.45 around 2124, more than 100 years from now. The upper bound of the 95\% CI reaches 0.45 in 2065, and the lower bound of the 95\% CI reaches 0.45 beyond the range of our projection. Appendix \ref{app:sensitivity} provides further discussion on model choice and the sensitivity of ARIMA projections to the choice of the equilibrium parameter.

We also make the somewhat concerning observation that the rate of growth in female author proportion has slowed in recent years, visible in Figures \ref{fig:cs_trends} and \ref{fig:prop_by_fos}. Our projection makes the optimistic assumption that the proportion will continue to grow towards or beyond parity, but the data may suggest otherwise. It remains to be seen whether a new trend is emerging that exhibits not an increase, but rather a leveling off or decrease in the proportion of female authors.

\subsection{Association of gender and co-authorship}

\begin{figure}[!tb]
\begin{center}
    \includegraphics[width=0.48\textwidth,keepaspectratio]{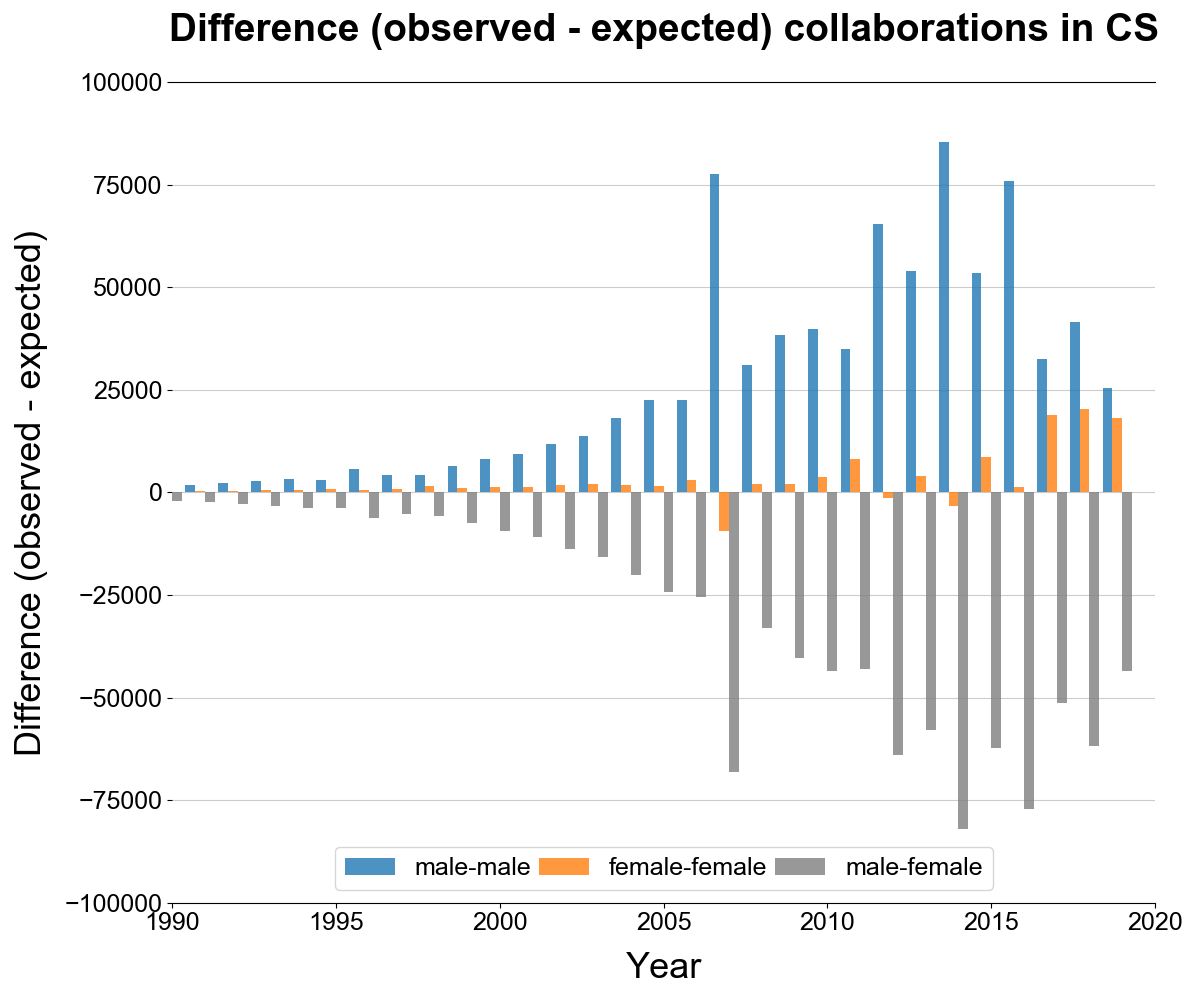}
    \hspace{4mm}
    \includegraphics[width=0.48\textwidth,keepaspectratio]{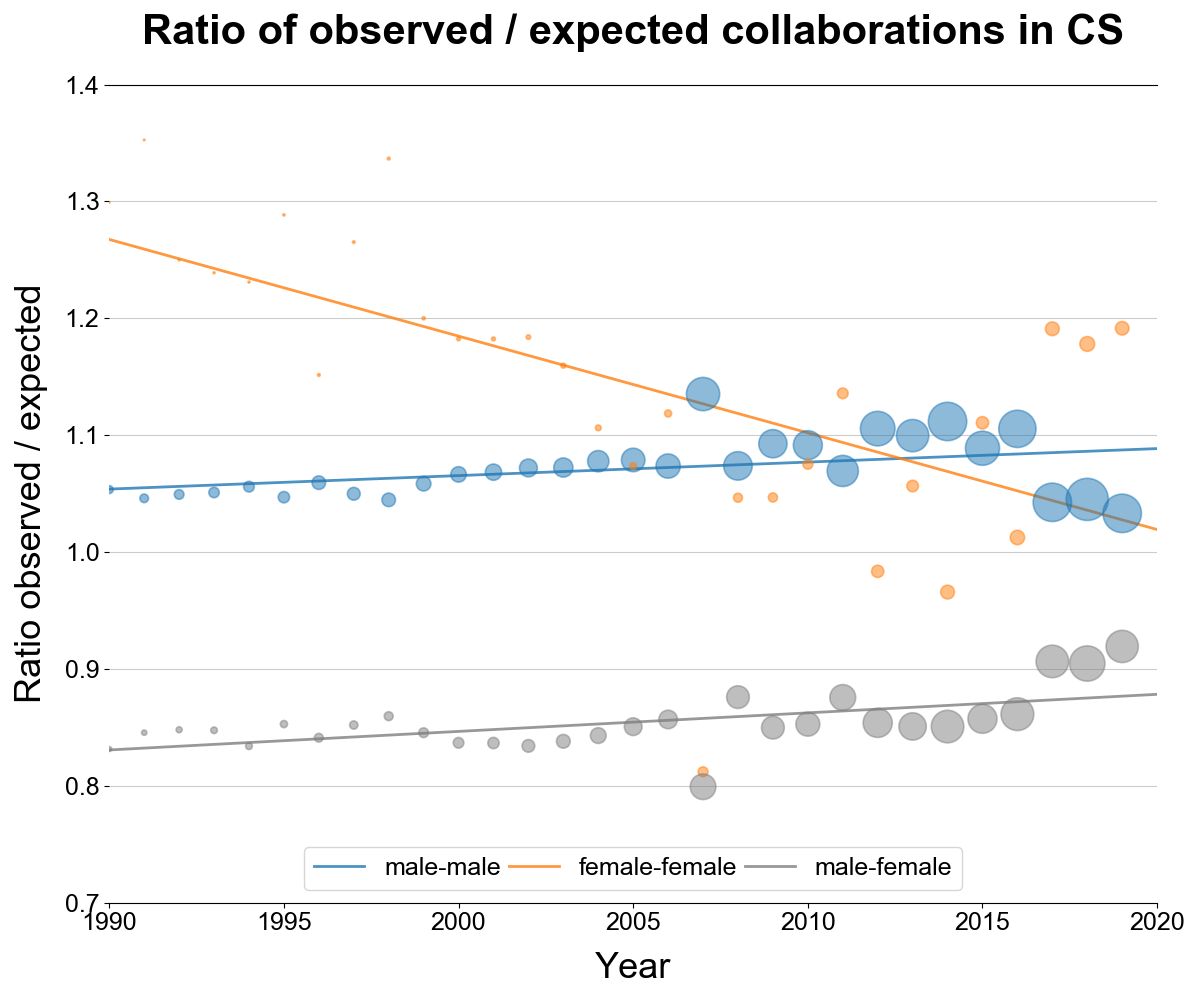}
	\caption{\label{fig:cs_coauthorship} The difference (\emph{left}) and ratio (\emph{right}) between observed and expected same- and cross-gender co-authorships in Computer Science since 1990. Marker size for the O/E ratio is proportional to the number of expected collaborations of that type in each year.}
\end{center}
\end{figure}

The numbers of same- (\emph{male-male} or \emph{female-female}) and cross-gender (\emph{male-female}) co-authorships in Computer Science are computed for each year. Figure \ref{fig:cs_coauthorship} shows the difference between the number of observed and expected collaborations of each type since 1990.\footnote{We show collaboration counts after 1990 because there is higher data volume in this period of time.} In this time period, there are more same-gender co-authorships than would be expected, and fewer cross-gender co-authorships than would be expected. In recent years, around 50000 cross-gender co-authorships per year were missing when compared to expected numbers.

The observed to expected ratio shows both optimistic and pessimistic collaboration trends. Although both men and women are more likely to co-author with authors of their own gender (positive O/E), the degree of same-gender bias is declining among female authors but potentially increasing among male authors. At the same time, the cross-gender collaboration gap (O/E $<$ 1.0) is still rather large, such that in recent years, only around 90\% of expected cross-gender collaborations are observed. In other words, although there are more opportunities for cross-gender collaboration in recent years (due to an increase in the number of female scientists working in the field), the observed number of cross-gender collaborations is still below what would be expected. Optimistically, these trends may be shifting in the recent past, with numbers from the last three years showing a shift towards more cross-gender co-authorship; although it is too early to say whether this tendency will preserve itself in the future.

\subsection{Comparison of CS with other fields of study}

\begin{figure}[!htb]
\includegraphics[width=0.9\textwidth,keepaspectratio]{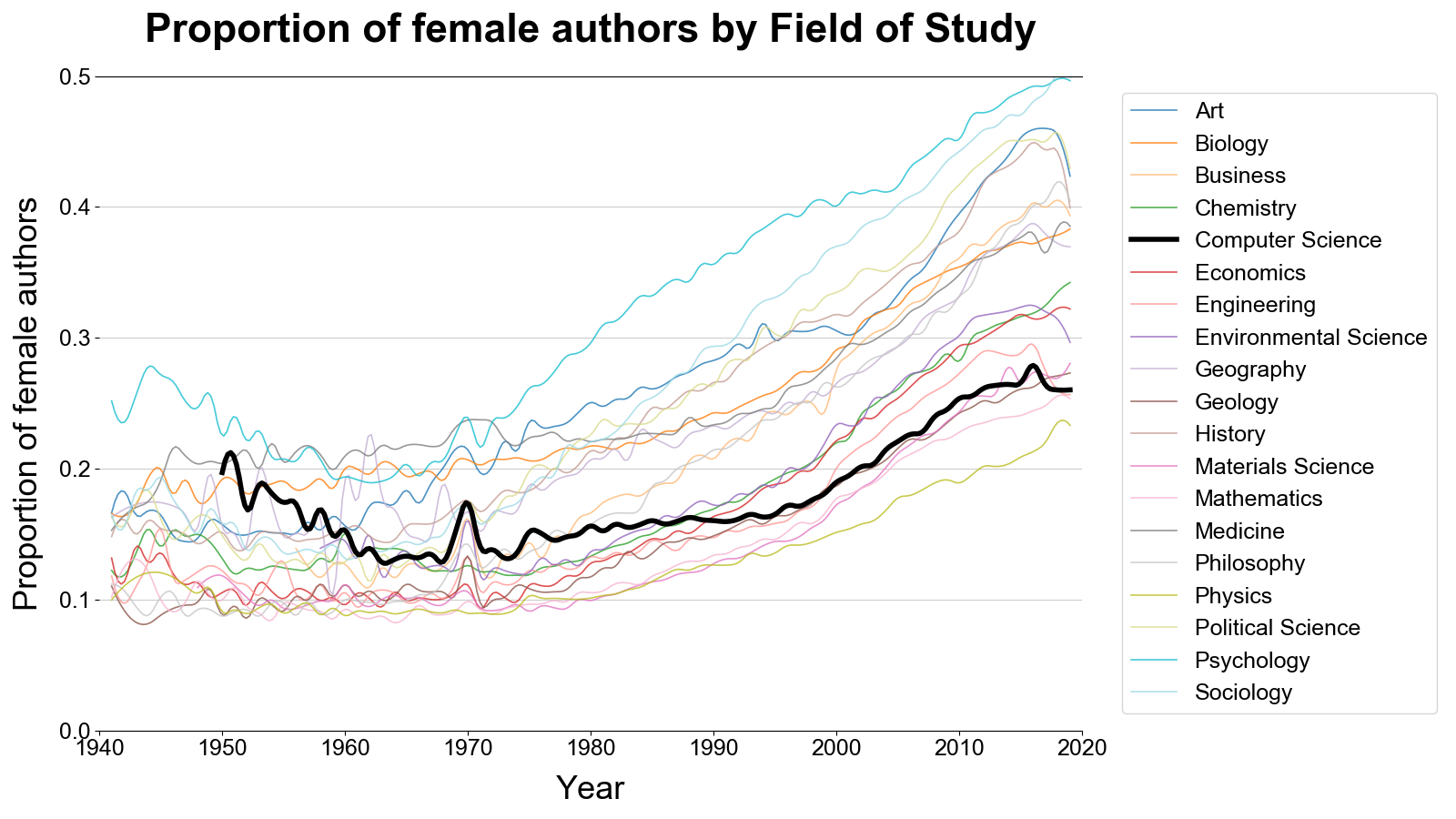}
\caption{\label{fig:prop_by_fos} The proportion of female authors among 19 fields of study. Proportion is plotted if there are more than 1000 author-paper units for which we could obtain gender information in a particular year.}
\end{figure}

Figure \ref{fig:prop_by_fos} shows the the proportion of female authors in 19 fields of study over the last 80 years. Computer Science is among the fields with the lowest female representation in recent years, despite having relatively higher female representation in the middle of the 20\textsuperscript{th} century. 

\section{Discussion}

Our analysis of the Computer Science literature reveals persistent patterns of inequality in gender and academic authorship. Although gender balance among authors is improving, progress is slower than we had hoped.

\subsection{Limitations}
\label{sec:limitations}

Inferring gender from first names is imperfect, and all gender-inference tools are subject to biases. Several studies have described and measured the differences between these services \citep{Karimi2016InferringGF, Santamaria2018ComparisonAB}. Based on results in \citet{Santamaria2018ComparisonAB}, Gender API has the lowest overall error rate but was slightly biased toward under-representation of females in their evaluation, in other words, the number of women estimated may be slightly lower than in reality. However, this bias may be offset by our sampling bias, since the population of CS authors is unlikely to be an unbiased sample of the general population, or the population whose names were used to construct the database behind Gender API. We attempt to mitigate some of these biases by treating the perceived gender as a probability distribution. One way to compute a more precise estimate is to weight the probabilities assigned by Gender API to each name using the prior probabilities of being a female or male CS author; this would likely produce a more pessimistic projection.

The proportion of authors with high uncertainty in Gender API results has also grown in our corpus over time. 
The average confidence of our gender predictions decreased from around 95\% in 1970-2000 to 90\% since 2005. We show and discuss this change in confidence in Appendix \ref{app:gender_api_uncertainty}.
While Gender API's average prediction confidence in our corpus is still high, this trend may pose a challenge for similar analysis in the future. 
Upon inspection of the data, we attribute this to the growing number of East Asian authors publishing in recent years. East Asian first names, when romanized, are more gender ambiguous. Gender API outperforms other gender lookup services, but still has lower overall confidence on names of East Asian origin \citep{Santamaria2018ComparisonAB}. In \citet{Mattauch2020ABA}, the authors explicitly exclude all authors with East Asian names from their name list during analysis, yet this accounts for the removal of more than 35\% of their dataset. Rather than remove an entire group of authors from our data, we believe that representing each author name as a distribution of gender probabilities offsets some of the issues of increasing gender ambiguity in our corpus over time.

We also recognize the limitations of using author-paper pairs as our units of measure. We do not distinguish between a person who is a single author on a paper, and a person who co-authors with many others. This biases our data by over-weighting authors in papers with more authors. Similarly, in our analysis of collaboration, we take each combination of authors for a paper as a collaborating pair, which again over-weights papers with more authors. In the Computer Science corpus, we observe an increase in the average authors per paper over time, growing to approximately 3.0 authors per paper in the last two years. However, Computer Science papers are still generally authored by smaller groups of individuals in the lower single digits, and we believe the bias introduced by our usage of author-paper pairs or collaborating author pairs to be minimal. 

Each author on a publication is also weighted equivalently in our analysis. We acknowledge that this discounts the special recognition extended to first authors, last authors, and single authors; we point readers to previous studies that have already demonstrated the distinctions between these groups \citep{West2013TheRO}.

Lastly, our projection of female author proportion uses data from the last 50 years to project more than 100 years into the future. We understand the inaccuracies of making such an extensive forecast with limited data. The goal of our projection is not to provide a definitive answer to the question of when gender parity will be reached among Computer Science authors; rather, the projection signals that even under optimistic growth, the gender gap will likely not close in the near future without some form of community or external intervention. Observed recent trends also suggest that the increase in female representation among Computer Science authors may be slowing in the last five years. The long range forecasts we show may not adequately capture changes on this shorter time scale. Our forecasts also do not reflect changes that would result from newly introduced or as yet unimplemented interventions.

\subsection{Prior work}

Inequality in gender representation is a well documented and studied issue in academia. Studies have shown that existent and perceived gender biases may affect many aspects of career and academic success, including but not limited to a woman's choice of college major \citep{Robnett2015GenderBI}, crediting in scientific publications \citep{Feldon2017TimetoCreditGI}, access to mentorship \citep{Decastro2014MentoringAT, Schluter2018TheGC, MossRacusin2012ScienceFS}, rate of promotions \citep{Clifton2019MathematicalMO}, opportunities for collaboration \citep{Elsevier2017Gender}, as well as publishing and citation trends \citep{Mattauch2020ABA, mohammad-2020-gender}. All of these factors can lead to imbalanced representation of women in certain fields of study.

With the increasing digitization of scholarly communication and availability of publication-related metadata, scholars have been able to better quantify inequality in authorship. \citet{Cohoon2011GenderAC} analyzed 86,000 ACM conference papers and showed increasing representation of women authors publishing at Computer Science venues, which strongly correlated with increasing numbers of female Computer Science PhDs \citep{Cohoon2011GenderAC}. \citet{West2013TheRO} analyzed 1.8 million papers from JSTOR, a large multi-disciplinary repository of academic literature, and revealed that although gender gaps are shrinking in academic publications, women were found to be significantly underrepresented as last and single authors. Elsevier, a large publisher of research articles, in an analysis of data from Scopus and ScienceDirect, reported the presence of gender imbalance among authors and inconsistent trends towards equal representation in different fields \citep{Elsevier2017Gender}. A study in 2018 confirmed continuing gender disparities among Nature Index journals, commonly considered some of the most reputable sources of academic literature, and in particular, limited representation of women among last authors, who are often perceived as more senior \citep{Bendels2018GenderDI}. Our work demonstrates that the gender gap is persistent and relatively large among Computer Science authors, which is consistent with the results of these studies.

A study of gender bias in authorship conducted by \citet{Holman2018TheGG} projected the closing of the gender gap in various fields based on recent trends. Through analyzing 9.1 million articles from PubMed, the authors projected that gender parity would be reached in around 20 years in certain biomedical fields such as Molecular Biology, Medicine, or Biochemistry. Holman et al.'s analysis of a small corpus of Computer Science pre-prints from arXiv showed that gender parity in Computer Science will be reached in more than 100 years from the present \citep{Holman2018TheGG}. Also corroborating our estimate is related work from \citet{Way2016GenderPA}, which forecasts that gender parity in CS faculty hiring will be reached around 2075. Due to the long duration of faculty careers, parity in hiring would be expected to precede parity in publication and overall representation. Our results confirm and expand upon the results of this prior work. We use a significantly larger corpus of literature metadata to place the trends observed in Computer Science in the context of other fields of study. Additionally, we provide an assessment of co-authorship trends, which demonstrate a gap in cross-gender collaborations among CS authors.

Major strides have been made to reduce gender disparities. The presence of an overall structure of sexism in academia continues to be debated \citep{Lundine2018TheGS, Boynton2018GenderBI, Lundine2019TheGS}, but many academic institutions recognize the issue and have sought to equalize admissions and hiring procedures. Evidence of movement towards more equitable representation in hiring and publication has been observed in some controlled settings \citep{Williams2015NationalHE, Hengel2017PublishingWF, Ceci:2011d9e}. How these observations translate into systemic change remain to be seen. Our results suggest, however, that the current pace of change in Computer Science will not result in a rapid closing of the gender gap.

\section{Conclusions}

We performed a large-scale analysis of the Computer Science literature (11.8M papers) to evaluate gender trends among authors. Based on trends over the last 50 years, the proportion of female authors in Computer Science is forecast to reach parity beyond the end of this century, and under different assumptions---it may take far longer. In this regard, Computer Science trails other fields of study, to which we may want to look for inspiration. We also observed lower than expected numbers of cross-gender collaborations, with a gap of approximately 50000 cross-gender collaborations per year in the last several years. 

Unless a major shift occurs that changes the gender makeup of the Computer Science community, the authorship gender gap will likely persist for a long time. Given the pervasiveness of computing technologies in our daily lives, it is of utmost importance that the designers and builders of these technologies reflect the diversity of their users. Gender is one type of diversity among many that can be more easily assessed using the types of automated methods we employ. We hope that these findings will motivate members of the community to reflect upon the causes of these disparities, and provide evidence to back up policy decisions to change the status quo.

\begin{acks}
  We would like to thank Jonathan Borchardt, Matt Gardner, and Candace Ross for conducting the initial analysis that motivated this work. We would also like to thank Kyle Lo for methodological discussions and Ashish Sabharwal, Maarten Sap, Noah Smith, and Mark Yatskar for helpful comments on earlier drafts of this paper.
\end{acks}

\bibliographystyle{ACM-Reference-Format}
\bibliography{references}


\begin{thebibliography}{28}

%%% the Web address.

\ifx \showCODEN    \undefined \def \showCODEN     #1{\unskip}     \fi
\ifx \showDOI      \undefined \def \showDOI       #1{#1}\fi
\ifx \showISBNx    \undefined \def \showISBNx     #1{\unskip}     \fi
\ifx \showISBNxiii \undefined \def \showISBNxiii  #1{\unskip}     \fi
\ifx \showISSN     \undefined \def \showISSN      #1{\unskip}     \fi
\ifx \showLCCN     \undefined \def \showLCCN      #1{\unskip}     \fi
\ifx \shownote     \undefined \def \shownote      #1{#1}          \fi
\ifx \showarticletitle \undefined \def \showarticletitle #1{#1}   \fi
\ifx \showURL      \undefined \def \showURL       {\relax}        \fi
% invisible to TeX
\providecommand\bibfield[2]{#2}
\providecommand\bibinfo[2]{#2}
\providecommand\natexlab[1]{#1}
\providecommand\showeprint[2][]{arXiv:#2}

\bibitem[\protect\citeauthoryear{??}{Els}{2017}]%
        {Elsevier2017Gender}
 \bibinfo{year}{2017}\natexlab{}.
\newblock \bibinfo{booktitle}{\emph{Gender in the global research landscape}}.
\newblock \bibinfo{type}{{T}echnical {R}eport}.
  \bibinfo{institution}{Elsevier}.
\newblock


\bibitem[\protect\citeauthoryear{Ammar, Groeneveld, Bhagavatula, Beltagy,
  Crawford, Downey, Dunkelberger, Elgohary, Feldman, Ha, Kinney, Kohlmeier, Lo,
  Murray, Ooi, Peters, Power, Skjonsberg, Wang, Wilhelm, Yuan, van Zuylen, and
  Etzioni}{Ammar et~al\mbox{.}}{2018}]%
        {Ammar2018ConstructionOT}
\bibfield{author}{\bibinfo{person}{Waleed Ammar}, \bibinfo{person}{Dirk
  Groeneveld}, \bibinfo{person}{Chandra Bhagavatula}, \bibinfo{person}{Iz
  Beltagy}, \bibinfo{person}{Miles Crawford}, \bibinfo{person}{Doug Downey},
  \bibinfo{person}{Jason Dunkelberger}, \bibinfo{person}{Ahmed Elgohary},
  \bibinfo{person}{Sergey Feldman}, \bibinfo{person}{Vu Ha},
  \bibinfo{person}{Rodney~Michael Kinney}, \bibinfo{person}{Sebastian
  Kohlmeier}, \bibinfo{person}{Kyle Lo}, \bibinfo{person}{Tyler~C. Murray},
  \bibinfo{person}{Hsu-Han Ooi}, \bibinfo{person}{Matthew~E. Peters},
  \bibinfo{person}{Joanna~L. Power}, \bibinfo{person}{Sam Skjonsberg},
  \bibinfo{person}{Lucy~Lu Wang}, \bibinfo{person}{Christopher Wilhelm},
  \bibinfo{person}{Zheng Yuan}, \bibinfo{person}{Madeleine van Zuylen}, {and}
  \bibinfo{person}{Oren Etzioni}.} \bibinfo{year}{2018}\natexlab{}.
\newblock \showarticletitle{Construction of the Literature Graph in Semantic
  Scholar}. In \bibinfo{booktitle}{\emph{NAACL-HLT}}.
\newblock


\bibitem[\protect\citeauthoryear{Bendels, Mueller, Brueggmann, and
  Groneberg}{Bendels et~al\mbox{.}}{2018}]%
        {Bendels2018GenderDI}
\bibfield{author}{\bibinfo{person}{Michael H.~K. Bendels},
  \bibinfo{person}{Ruth Mueller}, \bibinfo{person}{Doerthe Brueggmann}, {and}
  \bibinfo{person}{David~Alexander Groneberg}.}
  \bibinfo{year}{2018}\natexlab{}.
\newblock \showarticletitle{Gender disparities in high-quality research
  revealed by Nature Index journals}.
\newblock \bibinfo{journal}{\emph{PloS one}} \bibinfo{volume}{13},
  \bibinfo{number}{1} (\bibinfo{year}{2018}), \bibinfo{pages}{e0189136}.
\newblock
\urldef\tempurl%
\url{https://doi.org/10.1371/journal.pone.0189136}
\showDOI{\tempurl}


\bibitem[\protect\citeauthoryear{Box, Jenkins, and Reinsel}{Box
  et~al\mbox{.}}{1994}]%
        {TimeS1994Box}
\bibfield{author}{\bibinfo{person}{G.~E.~P. Box}, \bibinfo{person}{G.~M.
  Jenkins}, {and} \bibinfo{person}{G.~C. Reinsel}.}
  \bibinfo{year}{1994}\natexlab{}.
\newblock \bibinfo{booktitle}{\emph{Time series analysis: Forecasting and
  control} (\bibinfo{edition}{3} ed.)}.
\newblock \bibinfo{publisher}{Prentice Hall}, \bibinfo{address}{Englewood
  Cliffs, N.J.}
\newblock


\bibitem[\protect\citeauthoryear{Boynton, Georgiou, Reid, and Govus}{Boynton
  et~al\mbox{.}}{2018}]%
        {Boynton2018GenderBI}
\bibfield{author}{\bibinfo{person}{Jason~R Boynton}, \bibinfo{person}{Kristina
  Georgiou}, \bibinfo{person}{Mark Reid}, {and} \bibinfo{person}{Andrew
  Govus}.} \bibinfo{year}{2018}\natexlab{}.
\newblock \showarticletitle{Gender bias in publishing}.
\newblock \bibinfo{journal}{\emph{The Lancet}} \bibinfo{volume}{392},
  \bibinfo{number}{10157} (\bibinfo{year}{2018}), \bibinfo{pages}{1514--5}.
\newblock
\urldef\tempurl%
\url{https://doi.org/10.1016/S0140-6736(18)32000-2}
\showDOI{\tempurl}


\bibitem[\protect\citeauthoryear{Ceci and Williams}{Ceci and Williams}{2011}]%
        {Ceci:2011d9e}
\bibfield{author}{\bibinfo{person}{Stephen~J. Ceci} {and}
  \bibinfo{person}{Wendy~M. Williams}.} \bibinfo{year}{2011}\natexlab{}.
\newblock \showarticletitle{{Understanding current causes of women's
  underrepresentation in science}}.
\newblock \bibinfo{journal}{\emph{Proceedings of the National Academy of
  Sciences}} \bibinfo{volume}{108}, \bibinfo{number}{8} (\bibinfo{year}{2011}),
  \bibinfo{pages}{3157--3162}.
\newblock
\urldef\tempurl%
\url{https://doi.org/10.1073/pnas.1014871108}
\showDOI{\tempurl}


\bibitem[\protect\citeauthoryear{Clifton, Hill, Karamchandani, Autry, McMahon,
  and Sun}{Clifton et~al\mbox{.}}{2019}]%
        {Clifton2019MathematicalMO}
\bibfield{author}{\bibinfo{person}{Sara~M. Clifton}, \bibinfo{person}{Kaitlin
  Hill}, \bibinfo{person}{Avinash~J. Karamchandani}, \bibinfo{person}{Eric~A.
  Autry}, \bibinfo{person}{Patrick~J. McMahon}, {and} \bibinfo{person}{Grace
  Sun}.} \bibinfo{year}{2019}\natexlab{}.
\newblock \showarticletitle{Mathematical model of gender bias and homophily in
  professional hierarchies}.
\newblock \bibinfo{journal}{\emph{Chaos}}  \bibinfo{volume}{29}
  (\bibinfo{year}{2019}), \bibinfo{pages}{023135}.
\newblock
Issue 2.
\urldef\tempurl%
\url{https://doi.org/10.1063/1.5066450}
\showDOI{\tempurl}


\bibitem[\protect\citeauthoryear{Cohoon, Nigai, and Kaye}{Cohoon
  et~al\mbox{.}}{2011}]%
        {Cohoon2011GenderAC}
\bibfield{author}{\bibinfo{person}{Joanne~McGrath Cohoon},
  \bibinfo{person}{Sergey Nigai}, {and} \bibinfo{person}{Joseph Kaye}.}
  \bibinfo{year}{2011}\natexlab{}.
\newblock \showarticletitle{Gender and computing conference papers}.
\newblock \bibinfo{journal}{\emph{Commun. ACM}}  \bibinfo{volume}{54}
  (\bibinfo{year}{2011}), \bibinfo{pages}{72--80}.
\newblock
\urldef\tempurl%
\url{https://doi.org/10.1145/1978542.1978561}
\showDOI{\tempurl}


\bibitem[\protect\citeauthoryear{Decastro, Griffith, Ubel, Stewart, and
  Jagsi}{Decastro et~al\mbox{.}}{2014}]%
        {Decastro2014MentoringAT}
\bibfield{author}{\bibinfo{person}{Rochelle Decastro}, \bibinfo{person}{Kent~A.
  Griffith}, \bibinfo{person}{Peter~Anthony Ubel}, \bibinfo{person}{Abigail~J.
  Stewart}, {and} \bibinfo{person}{Reshma Jagsi}.}
  \bibinfo{year}{2014}\natexlab{}.
\newblock \showarticletitle{Mentoring and the career satisfaction of male and
  female academic medical faculty}.
\newblock \bibinfo{journal}{\emph{Academic medicine : journal of the
  Association of American Medical Colleges}} \bibinfo{volume}{89},
  \bibinfo{number}{2} (\bibinfo{year}{2014}), \bibinfo{pages}{301--11}.
\newblock
\urldef\tempurl%
\url{https://doi.org/10.1097/ACM.0000000000000109}
\showDOI{\tempurl}


\bibitem[\protect\citeauthoryear{Feldon, Peugh, Maher, Roksa, and
  Tofel-Grehl}{Feldon et~al\mbox{.}}{2017}]%
        {Feldon2017TimetoCreditGI}
\bibfield{author}{\bibinfo{person}{David~F. Feldon}, \bibinfo{person}{James~L.
  Peugh}, \bibinfo{person}{Michelle~A. Maher}, \bibinfo{person}{Josipa Roksa},
  {and} \bibinfo{person}{Colby Tofel-Grehl}.} \bibinfo{year}{2017}\natexlab{}.
\newblock \showarticletitle{Time-to-Credit Gender Inequities of First-Year
  {PhD} Students in the Biological Sciences}.
\newblock \bibinfo{journal}{\emph{CBE life sciences education}}
  \bibinfo{volume}{16} (\bibinfo{year}{2017}), \bibinfo{pages}{ar4}.
\newblock
Issue 1.
\urldef\tempurl%
\url{https://doi.org/10.1187/cbe.16-08-0237}
\showDOI{\tempurl}


\bibitem[\protect\citeauthoryear{Fernandes and Monteiro}{Fernandes and
  Monteiro}{2016}]%
        {Fernandes2016EvolutionIT}
\bibfield{author}{\bibinfo{person}{Jo{\~a}o~M. Fernandes} {and}
  \bibinfo{person}{Miguel~Pessoa Monteiro}.} \bibinfo{year}{2016}\natexlab{}.
\newblock \showarticletitle{Evolution in the number of authors of computer
  science publications}.
\newblock \bibinfo{journal}{\emph{Scientometrics}}  \bibinfo{volume}{110}
  (\bibinfo{year}{2016}), \bibinfo{pages}{529--539}.
\newblock


\bibitem[\protect\citeauthoryear{Hengel}{Hengel}{2017}]%
        {Hengel2017PublishingWF}
\bibfield{author}{\bibinfo{person}{Erin Hengel}.}
  \bibinfo{year}{2017}\natexlab{}.
\newblock \showarticletitle{Publishing while Female. Are women held to higher
  standards? {E}vidence from peer review.}
\newblock \bibinfo{journal}{\emph{Cambridge Working Paper Economics}}
  \bibinfo{volume}{1753} (\bibinfo{year}{2017}).
\newblock
\urldef\tempurl%
\url{https://doi.org/10.17863/CAM.17548}
\showDOI{\tempurl}


\bibitem[\protect\citeauthoryear{Holman, Stuart-Fox, and Hauser}{Holman
  et~al\mbox{.}}{2018}]%
        {Holman2018TheGG}
\bibfield{author}{\bibinfo{person}{Luke Holman}, \bibinfo{person}{Devi
  Stuart-Fox}, {and} \bibinfo{person}{Cindy~E. Hauser}.}
  \bibinfo{year}{2018}\natexlab{}.
\newblock \showarticletitle{The gender gap in science: How long until women are
  equally represented?}
\newblock \bibinfo{journal}{\emph{PLoS biology}} \bibinfo{volume}{16},
  \bibinfo{number}{4} (\bibinfo{year}{2018}), \bibinfo{pages}{e2004956}.
\newblock
\urldef\tempurl%
\url{https://doi.org/10.1371/journal.pbio.2004956}
\showDOI{\tempurl}


\bibitem[\protect\citeauthoryear{Hyndman and Khandakar}{Hyndman and
  Khandakar}{2008}]%
        {Automatic2008Hyndman}
\bibfield{author}{\bibinfo{person}{Rob~J Hyndman} {and}
  \bibinfo{person}{Yeasmin Khandakar}.} \bibinfo{year}{2008}\natexlab{}.
\newblock \showarticletitle{Automatic time series forecasting: the forecast
  package for {R}}.
\newblock \bibinfo{journal}{\emph{Journal of Statistical Software}}
  \bibinfo{volume}{26}, \bibinfo{number}{3} (\bibinfo{year}{2008}),
  \bibinfo{pages}{1--22}.
\newblock
\urldef\tempurl%
\url{https://doi.org/10.18637/jss.v027.i03}
\showDOI{\tempurl}


\bibitem[\protect\citeauthoryear{Karimi, Wagner, Lemmerich, Jadidi, and
  Strohmaier}{Karimi et~al\mbox{.}}{2016}]%
        {Karimi2016InferringGF}
\bibfield{author}{\bibinfo{person}{Fariba Karimi}, \bibinfo{person}{Claudia
  Wagner}, \bibinfo{person}{Florian Lemmerich}, \bibinfo{person}{Mohsen
  Jadidi}, {and} \bibinfo{person}{Markus Strohmaier}.}
  \bibinfo{year}{2016}\natexlab{}.
\newblock \showarticletitle{Inferring Gender from Names on the Web: A
  Comparative Evaluation of Gender Detection Methods}. In
  \bibinfo{booktitle}{\emph{WWW}}.
\newblock
\urldef\tempurl%
\url{https://doi.org/10.1145/2872518.2889385}
\showDOI{\tempurl}


\bibitem[\protect\citeauthoryear{Lundine, Bourgeault, Clark, Heidari, and
  Balabanova}{Lundine et~al\mbox{.}}{2018}]%
        {Lundine2018TheGS}
\bibfield{author}{\bibinfo{person}{Jamie Lundine}, \bibinfo{person}{Ivy~Lynn
  Bourgeault}, \bibinfo{person}{Jocalyn Clark}, \bibinfo{person}{Shirin
  Heidari}, {and} \bibinfo{person}{Dina Balabanova}.}
  \bibinfo{year}{2018}\natexlab{}.
\newblock \showarticletitle{The gendered system of academic publishing}.
\newblock \bibinfo{journal}{\emph{The Lancet}} \bibinfo{volume}{391},
  \bibinfo{number}{10132} (\bibinfo{year}{2018}), \bibinfo{pages}{1754--6}.
\newblock
\urldef\tempurl%
\url{https://doi.org/10.1016/S0140-6736(18)30950-4}
\showDOI{\tempurl}


\bibitem[\protect\citeauthoryear{Lundine, Bourgeault, Clark, Heidari, and
  Balabanova}{Lundine et~al\mbox{.}}{2019}]%
        {Lundine2019TheGS}
\bibfield{author}{\bibinfo{person}{Jamie Lundine}, \bibinfo{person}{Ivy~Lynn
  Bourgeault}, \bibinfo{person}{Jocalyn Clark}, \bibinfo{person}{Shirin
  Heidari}, {and} \bibinfo{person}{Dina Balabanova}.}
  \bibinfo{year}{2019}\natexlab{}.
\newblock \showarticletitle{Gender bias in academia}.
\newblock \bibinfo{journal}{\emph{The Lancet}} \bibinfo{volume}{393},
  \bibinfo{number}{10173} (\bibinfo{year}{2019}), \bibinfo{pages}{741--3}.
\newblock
\urldef\tempurl%
\url{https://doi.org/10.1016/S0140-6736(19)30281-8}
\showDOI{\tempurl}


\bibitem[\protect\citeauthoryear{Mattauch, Lohmann, Hannig, Lohmann, and
  Teich}{Mattauch et~al\mbox{.}}{2020}]%
        {Mattauch2020ABA}
\bibfield{author}{\bibinfo{person}{Sandra Mattauch}, \bibinfo{person}{K.
  Lohmann}, \bibinfo{person}{F. Hannig}, \bibinfo{person}{Daniel Lohmann},
  {and} \bibinfo{person}{J. Teich}.} \bibinfo{year}{2020}\natexlab{}.
\newblock \showarticletitle{A bibliometric approach for detecting the gender
  gap in computer science}.
\newblock \bibinfo{journal}{\emph{Commun. ACM}}  \bibinfo{volume}{63}
  (\bibinfo{year}{2020}), \bibinfo{pages}{74--80}.
\newblock


\bibitem[\protect\citeauthoryear{Mohammad}{Mohammad}{2020}]%
        {mohammad-2020-gender}
\bibfield{author}{\bibinfo{person}{Saif~M. Mohammad}.}
  \bibinfo{year}{2020}\natexlab{}.
\newblock \showarticletitle{Gender Gap in Natural Language Processing Research:
  Disparities in Authorship and Citations}. In
  \bibinfo{booktitle}{\emph{Proceedings of the 58th Annual Meeting of the
  Association for Computational Linguistics}}. \bibinfo{publisher}{Association
  for Computational Linguistics}, \bibinfo{address}{Online},
  \bibinfo{pages}{7860--7870}.
\newblock
\urldef\tempurl%
\url{https://doi.org/10.18653/v1/2020.acl-main.702}
\showDOI{\tempurl}


\bibitem[\protect\citeauthoryear{Moss-Racusin, Dovidio, Brescoll, Graham, and
  Handelsman}{Moss-Racusin et~al\mbox{.}}{2012}]%
        {MossRacusin2012ScienceFS}
\bibfield{author}{\bibinfo{person}{Corinne~A. Moss-Racusin},
  \bibinfo{person}{John~F. Dovidio}, \bibinfo{person}{Victoria~L. Brescoll},
  \bibinfo{person}{Mark~J. Graham}, {and} \bibinfo{person}{Jo Handelsman}.}
  \bibinfo{year}{2012}\natexlab{}.
\newblock \showarticletitle{Science faculty's subtle gender biases favor male
  students}.
\newblock \bibinfo{journal}{\emph{Proceedings of the National Academy of
  Sciences of the United States of America}}  \bibinfo{volume}{109}
  (\bibinfo{year}{2012}), \bibinfo{pages}{16474--9}.
\newblock
Issue 41.
\urldef\tempurl%
\url{https://doi.org/10.1073/pnas.1211286109}
\showDOI{\tempurl}


\bibitem[\protect\citeauthoryear{Robnett}{Robnett}{2015}]%
        {Robnett2015GenderBI}
\bibfield{author}{\bibinfo{person}{Rachael~D. Robnett}.}
  \bibinfo{year}{2015}\natexlab{}.
\newblock \showarticletitle{Gender bias in {STEM} fields: variation in
  prevalence and links to {STEM} self-concept}.
\newblock \bibinfo{journal}{\emph{Psychology of Women Quarterly}}
  (\bibinfo{year}{2015}).
\newblock
\urldef\tempurl%
\url{https://doi.org/10.1177/0361684315596162}
\showDOI{\tempurl}


\bibitem[\protect\citeauthoryear{Santamar\'{i}a and
  Mihaljevi\'{c}}{Santamar\'{i}a and Mihaljevi\'{c}}{2018}]%
        {Santamaria2018ComparisonAB}
\bibfield{author}{\bibinfo{person}{Luc\'{i}a~Prieto Santamar\'{i}a} {and}
  \bibinfo{person}{Helena Mihaljevi\'{c}}.} \bibinfo{year}{2018}\natexlab{}.
\newblock \showarticletitle{Comparison and benchmark of name-to-gender
  inference services}.
\newblock \bibinfo{journal}{\emph{PeerJ Computer Science}}  \bibinfo{volume}{4}
  (\bibinfo{year}{2018}), \bibinfo{pages}{e156}.
\newblock
\urldef\tempurl%
\url{https://doi.org/10.7717/peerj-cs.156}
\showDOI{\tempurl}


\bibitem[\protect\citeauthoryear{Schluter}{Schluter}{2018}]%
        {Schluter2018TheGC}
\bibfield{author}{\bibinfo{person}{Natalie Schluter}.}
  \bibinfo{year}{2018}\natexlab{}.
\newblock \showarticletitle{The glass ceiling in {NLP}}. In
  \bibinfo{booktitle}{\emph{EMNLP}}.
\newblock
\urldef\tempurl%
\url{https://doi.org/10.18653/v1/D18-1301}
\showDOI{\tempurl}


\bibitem[\protect\citeauthoryear{Shapiro and Wilk}{Shapiro and Wilk}{1965}]%
        {AnA1965Shapiro}
\bibfield{author}{\bibinfo{person}{S.~S. Shapiro} {and} \bibinfo{person}{M.~B.
  Wilk}.} \bibinfo{year}{1965}\natexlab{}.
\newblock \showarticletitle{An analysis of variance test for normality
  (complete samples)}.
\newblock \bibinfo{journal}{\emph{Biometrika}}  \bibinfo{volume}{52}
  (\bibinfo{year}{1965}), \bibinfo{pages}{591--611}.
\newblock
Issue 3-4.
\urldef\tempurl%
\url{https://doi.org/10.1093/biomet/52.3-4.591}
\showDOI{\tempurl}


\bibitem[\protect\citeauthoryear{Shen, Ma, and Wang}{Shen
  et~al\mbox{.}}{2018}]%
        {shen-etal-2018-web}
\bibfield{author}{\bibinfo{person}{Zhihong Shen}, \bibinfo{person}{Hao Ma},
  {and} \bibinfo{person}{Kuansan Wang}.} \bibinfo{year}{2018}\natexlab{}.
\newblock \showarticletitle{A Web-scale system for scientific knowledge
  exploration}. In \bibinfo{booktitle}{\emph{Proceedings of {ACL} 2018, System
  Demonstrations}}. \bibinfo{publisher}{Association for Computational
  Linguistics}, \bibinfo{address}{Melbourne, Australia},
  \bibinfo{pages}{87--92}.
\newblock
\urldef\tempurl%
\url{https://doi.org/10.18653/v1/P18-4015}
\showDOI{\tempurl}


\bibitem[\protect\citeauthoryear{Way, Larremore, and Clauset}{Way
  et~al\mbox{.}}{2016}]%
        {Way2016GenderPA}
\bibfield{author}{\bibinfo{person}{Samuel~F. Way}, \bibinfo{person}{Daniel~B.
  Larremore}, {and} \bibinfo{person}{Aaron Clauset}.}
  \bibinfo{year}{2016}\natexlab{}.
\newblock \showarticletitle{Gender, productivity, and prestige in computer
  science faculty hiring networks}. In \bibinfo{booktitle}{\emph{WWW}}.
\newblock
\urldef\tempurl%
\url{https://doi.org/10.1145/2872427.2883073}
\showDOI{\tempurl}


\bibitem[\protect\citeauthoryear{West, Jacquet, King, Correll, and
  Bergstrom}{West et~al\mbox{.}}{2013}]%
        {West2013TheRO}
\bibfield{author}{\bibinfo{person}{Jevin~D. West}, \bibinfo{person}{Jennifer
  Jacquet}, \bibinfo{person}{Molly~M. King}, \bibinfo{person}{Shelley~J.
  Correll}, {and} \bibinfo{person}{Carl~T. Bergstrom}.}
  \bibinfo{year}{2013}\natexlab{}.
\newblock \showarticletitle{The Role of Gender in Scholarly Authorship}.
\newblock \bibinfo{journal}{\emph{PloS one}} \bibinfo{volume}{8},
  \bibinfo{number}{7} (\bibinfo{year}{2013}), \bibinfo{pages}{e66212}.
\newblock
\urldef\tempurl%
\url{https://doi.org/10.1371/journal.pone.0066212}
\showDOI{\tempurl}


\bibitem[\protect\citeauthoryear{Williams and Ceci}{Williams and Ceci}{2015}]%
        {Williams2015NationalHE}
\bibfield{author}{\bibinfo{person}{W.~Mattieu Williams} {and}
  \bibinfo{person}{Stephen~J Ceci}.} \bibinfo{year}{2015}\natexlab{}.
\newblock \showarticletitle{National hiring experiments reveal 2:1 faculty
  preference for women on STEM tenure track}.
\newblock \bibinfo{journal}{\emph{Proceedings of the National Academy of
  Sciences of the United States of America}} \bibinfo{volume}{112},
  \bibinfo{number}{17} (\bibinfo{year}{2015}), \bibinfo{pages}{5360--5}.
\newblock
\urldef\tempurl%
\url{https://doi.org/10.1073/pnas.1418878112}
\showDOI{\tempurl}


\end{thebibliography}

\pagebreak
\appendix

\section{Sensitivity analysis for parity projection}
\label{app:sensitivity}

\begin{figure}[!h]
\begin{center}
    \includegraphics[width=0.55\textwidth,keepaspectratio]{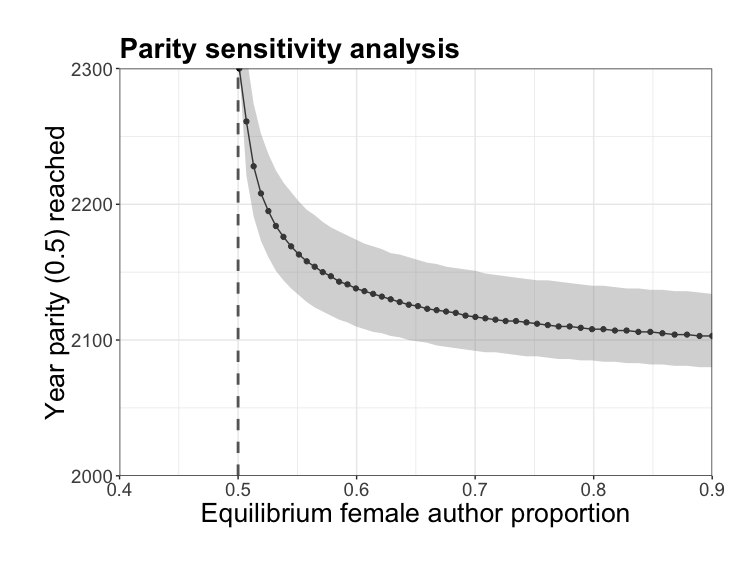}
	\caption{\label{fig:dblp_sensitivity} The equilibrium female author proportion parameter affects the year that parity is reached. The expected year for reaching exact parity (point at which proportion crosses 0.5) is shown along with 95\% confidence intervals.}
\end{center}
\end{figure}

Figure \ref{fig:dblp_sensitivity} shows a sensitivity analysis over the equilibrium female author proportion parameter $\alpha$. This analysis shows the year in which parity is first reached at each equilibrium proportion; note that when $\alpha=0.5$, exact 50/50 parity is, by definition, never attained in finite time. We therefore report the time at which the female author proportion surpasses 0.45, within 10\% of exact parity. When the equilibrium proportion is expected to favor women over men (above 0.5), the year in which parity is reached occurs earlier. Even with the aggressive projection that women will eventually author 90\% of all publications, the expected year in which 50/50 parity will be reached is still around 2100.

\section{Average authors per paper}
\label{app:app_over_time}

\begin{figure}[!h]
\begin{center}
    \includegraphics[width=0.9\textwidth,keepaspectratio]{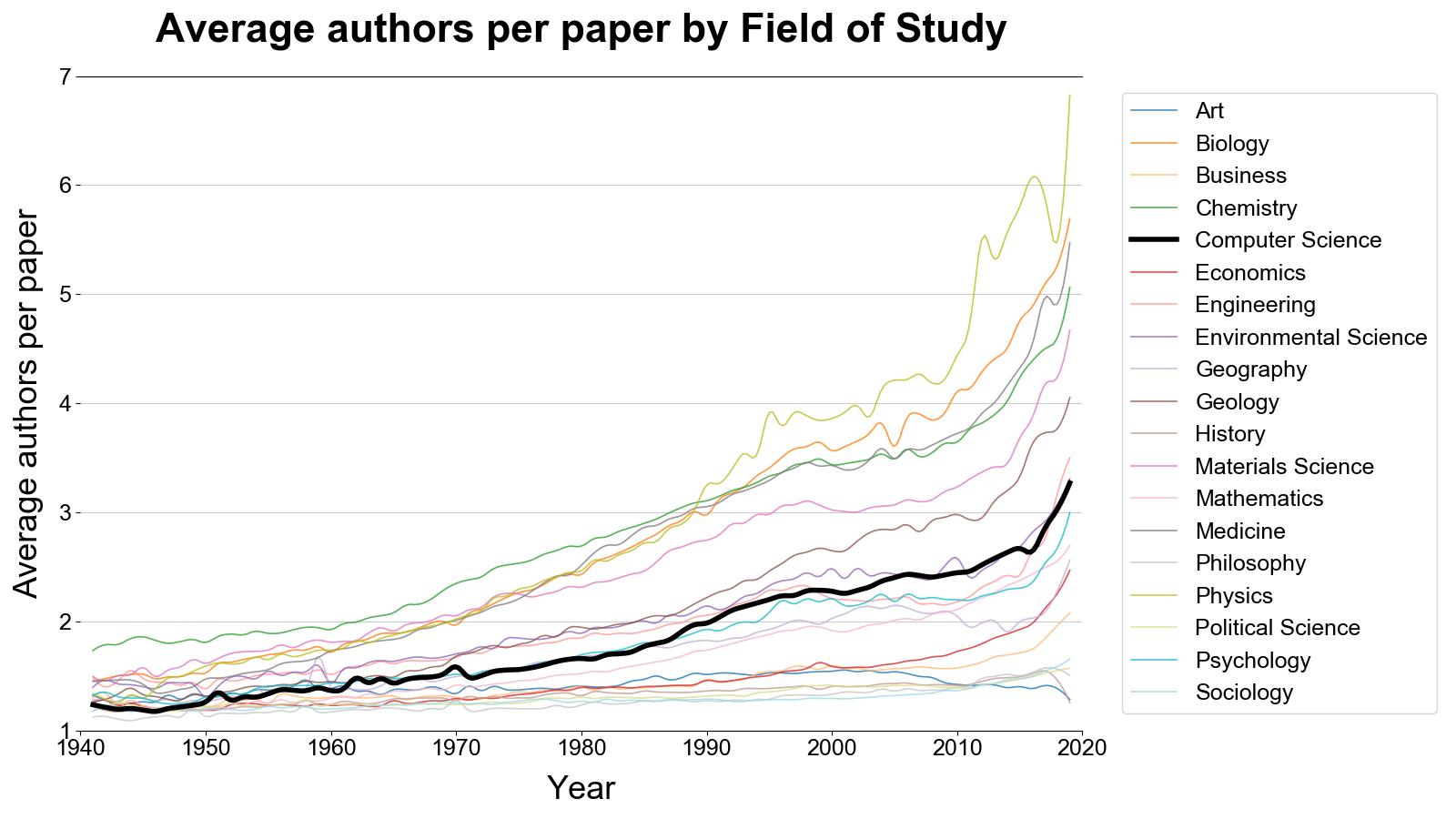}
	\caption{\label{fig:app_by_fos} Average authors per paper in different fields of study.}
\end{center}
\end{figure}

Figure \ref{fig:app_by_fos} shows the average number of authors per paper over the years. Numbers of authors per paper have increased significantly in recent years, especially in fields dominated by large-scale experiments, like Physics, Biology, and Medicine. Although Computer Science has also seen an increase in contributing authors per paper, this growth is much slower relative to other scientific fields. This stands in contrast to fields that are closer to the humanities, such as Art, History, or Philosophy.

\section{Assumption of logistic fit}
\label{app:fit_rmse}

We assume the growth of the proportion of female authors observes logistic behavior. The choice of logistic fit is natural because the proportion of female authors is necessarily bound by 0 and 1, which is a feature of logistic growth curves. In this Appendix, we further justify the choice of logistic fit by comparing it against linear and exponential fits, which could be better descriptors of rates of change in the short term. For each field of study, we fit a linear, exponential, and logistic curve to the proportion of female authors over time, beginning in 1970, when the trends primarily show consistent growth. We compute the RMSE of each fit and provide this in Table \ref{tab:fit_rmse}. For all fields of study, the RMSE of logistic fit is the lowest of the three curve types.

\begin{table*}[tb!]
    \small
    \centering
    \begin{tabular}{lccc}
        \toprule
         Field of Study & RMSE (linear) & RMSE (exponential) & RMSE (logistic) \\
        \midrule
         Art & 0.14 & 0.11 & 0.10 \\
         Biology & 0.09 & 0.07 & 0.02 \\
         Business & 0.09 & 0.08 & 0.05 \\
         Chemistry & 0.09 & 0.05 & 0.03 \\
         Computer Science & 0.10 & 0.08 & 0.04 \\
         Economics & 0.07 & 0.06 & 0.03 \\
         Engineering & 0.13 & 0.11 & 0.07 \\
         Environmental Science & 0.10 & 0.09 & 0.06 \\
         Geography & 0.10 & 0.08 & 0.07 \\
         Geology & 0.05 & 0.03 & 0.03 \\
         History & 0.08 & 0.09 & 0.07 \\
         Materials Science & 0.11 & 0.06 & 0.03 \\
         Mathematics & 0.06 & 0.04 & 0.02 \\
         Medicine & 0.14 & 0.11 & 0.03 \\
         Philosophy & 0.09 & 0.05 & 0.04 \\
         Physics & 0.06 & 0.03 & 0.02 \\
         Political Science & 0.08 & 0.09 & 0.07 \\
         Psychology & 0.07 & 0.11 & 0.05 \\
         Sociology & 0.03 & 0.12 & 0.02 \\
        \bottomrule
    \end{tabular}
    \caption{RMSE of different curve fits for the proportion of female authors in each field of study since 1970.}
    \label{tab:fit_rmse}
\end{table*}

\section{Uncertainty in author gender}
\label{app:gender_api_uncertainty}

\begin{figure}[!h]
\begin{center}
	\includegraphics[width=0.5\textwidth,keepaspectratio]{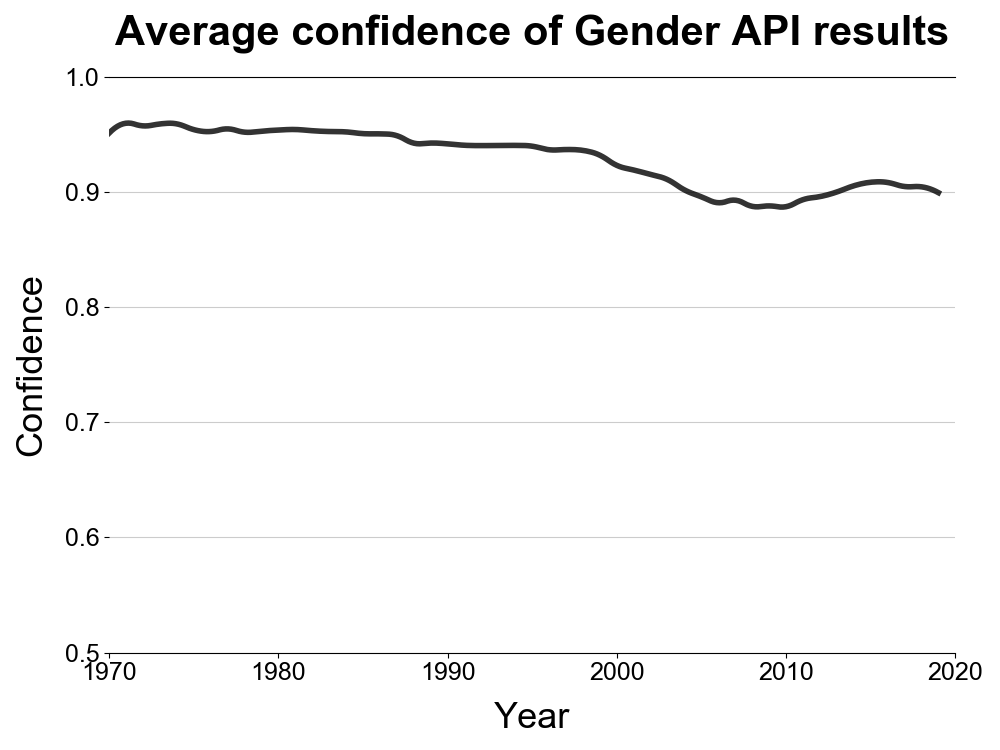}
	\caption{\label{fig:confidence} The average author gender confidence of Gender API on Computer Science authors, per publication year.}
\end{center}
\end{figure}

The average confidence of Gender API results averaged over the authors in each year is shown in Figure \ref{fig:confidence}. The average confidence of Gender API is decreasing over time in our corpus, and experienced a drop after 2000 to around 90\% in recent years. We speculate about the cause of this drop in \S \ref{sec:limitations}. Analyses such as ours may become increasingly difficult to perform in the future unless the datasets behind gendering services are improved.

\end{document}